\documentstyle[12pt,epsfig]{article}
\newcommand{\Dzero}{D\O{} }
\newcommand{\etal}{{\em et al. }}
\newcommand{\Dyrad}{D{\sc yrad} }
\newcommand{\slE}{\mbox{/}\!\!\!\!E}
\newcommand{\GeV}{\mbox{ GeV}}

\begin{document}
\begin{titlepage}
\vspace*{-1cm}
\begin{flushright}
CERN--TH/97--173\\
DTP/97/72 \\
July 1997 \\
\end{flushright}                                

\vskip 1.cm
\begin{center}                                                             

{\Large\bf
What does the $W$ transverse momentum distribution\\
 say about the $W+1$~jet/$W+0$~jet ratio?}
\vskip 1.3cm
{\large E.~W.~N.~Glover$^{a,b}$ and  D.~J.~Summers$^b$}
\vskip .2cm
$^a$~{\it Department of Physics, University of Durham,
Durham DH1 3LE, England }\\
$^b$~{\it TH Division, CERN, CH-1211 Geneva 23, Switzerland}
\vskip 2.3cm   

\end{center}      

\begin{abstract}
We compute the $W$ transverse momentum distribution at large
transverse momentum at fixed next-to-leading order (i.e. ${\cal
O}(\alpha_s^2)$) and find good agreement with the preliminary
measurements reported by \Dzero.  We find that the $W$ transverse
momentum distribution is typically significantly larger than the
exclusive one-jet rate for a given value of $p_T^W$ or $E_T^{\rm
jet}$, mainly because two or more jet events are excluded.  As a
consequence, we find that theoretically the $W+~1$ jet to $W+~0$ jet
ratio ${\cal R}^{10}$ is smaller than the analogous quantity defined
in terms of $p_T^W$, ${\cal R}^{W}$.  This hierarchy is preserved
under changes of renormalisation/factorisation scales, strong coupling
constant and jet algorithm.  However, this appears to be in
contradiction with the preliminary experimental results which suggest
${\cal R}^{10} > {\cal R}^{W}$.
\end{abstract}                                                                

\vfill
\begin{flushleft}
CERN--TH/97--173\\
July 1997
\end{flushleft}
\end{titlepage}
\newpage                                                                     

During the past year the \Dzero collaboration has reported 
preliminary measurements of the 
ratio ${\cal R}^{10}$ \cite{R10} defined by \cite{UA2,UA1},
\begin{equation}
{\cal R}^{10}(E_T^{\rm min}) 
={\sigma (W + \mbox{1 jet} ) \over  \sigma (W + \mbox{0 jet} )}\;,
\end{equation}
where the jets are defined with transverse energy above some $E_T^{\rm
min}$.  At present the data appears to indicate larger values of
${\cal R}^{10}$ than those obtained in next-to-leading order QCD
predictions for the same quantity \cite{R10old,Jaethesis,R10,GGK,GG}.  There have already been several
suggested explanations for this excess.  For example, Bal\'azs and
Yuan \cite{BY} have considered the effect of soft gluon resummation on
the related quantity ${\cal R}^{W}$,
\begin{equation}
{\cal R}^{W}(p_T^{W,\rm min}) 
={\sigma (W , p_T^W > p_T^{W,\rm min} ) \over
\sigma (W , p_T^W < p_T^{W,\rm min} )}\;.
\end{equation}
New physics effects are also possible and Choudhury \etal \cite{chou}
have considered the effect that a massive vector boson with the
quantum numbers of both a $W$ boson and a gluon would have on the
observed value of ${\cal R}^{10}$.  A more mundane explanation is that
an increase in the gluon parton distribution at medium Bjorken $x$
values would boost the $W$ +~1 jet rate, which receives contributions
from $qg$ scattering, while having little effect on the zero jet rate
\cite{R10}.  Here we discuss the relationship between the large
transverse momentum distribution of the $W$ boson, for which the
\Dzero collaboration recently reported a preliminary measurement
\cite{Wpt}, and the ${\cal R}^{10}$ ratio.  We also consider the
extent to which we can accurately theoretically predict each of these
quantities.

At leading order in perturbative QCD the large $p_T$
distribution of the $W$ boson is given by the $W$ recoiling against a
single parton. Typically, the parton will hadronize into an observed
jet, with the jet $E_T$ balancing the $W$ transverse momentum,
so that,
\begin{equation}
{\cal R}^{W}(p_T^{W,\rm min}) \simeq {\cal R}^{10}(E_T^{\rm min} 
= p_T^{W,\rm min}) \;.
\end{equation}
Beyond lowest order this relationship is no longer strictly true at 
the parton level,
however it makes the quantity ${\cal R}^{W}$ worth studying in the 
context of the measured
\Dzero excess in ${\cal R}^{10}$. At the moment \Dzero make no 
direct measurement
of the ${\cal R}^{W}$ ratio. However, for a different analysis
\cite{Wpt} they have made a preliminary measurement of the normalised
transverse momentum distribution of the $W$,$1/\sigma\, d\sigma /
dp_T^W$, and from this measurement we can construct ${\cal R}^{W}$,
\begin{equation}
{\cal R}^{W}(p_T^{W,\rm min}) = 
{\displaystyle   \int_{p_T^{W,\rm min}}^\infty d p_T^W {1\over\sigma}\, 
                                {d\sigma\over dp_T^W}   \over
1 - \displaystyle \int_{p_T^{W,\rm min}}^\infty dp_T^W   
                                  {1\over\sigma}\, 
                                {d\sigma\over dp_T^W} } \;.
\label{eq:ptwtorw}
\end{equation}

\begin{figure}
\begin{center}
\psfig{figure=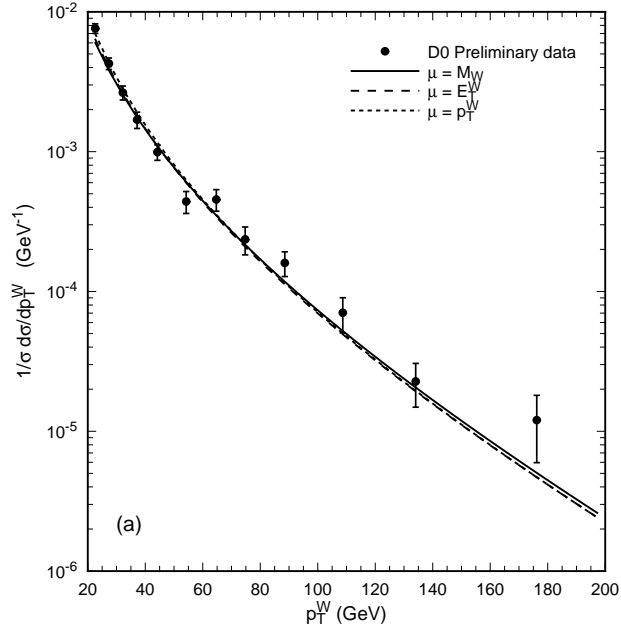,height=9cm}\\
\psfig{figure=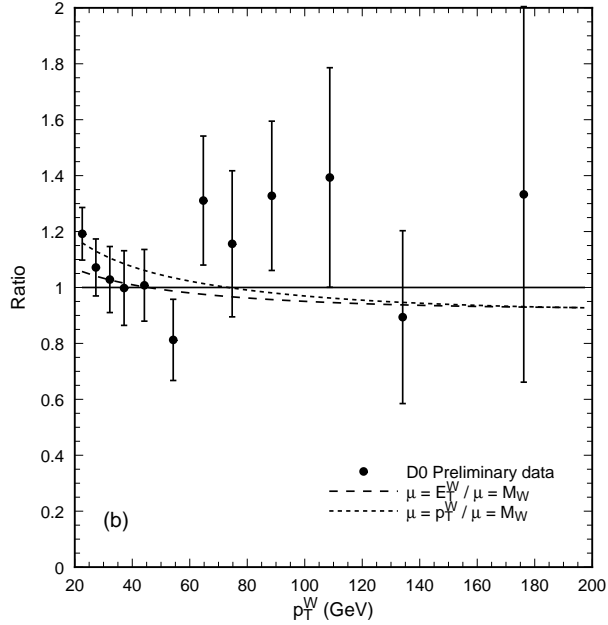,height=9cm}
\end{center}
\caption{a) The \Dzero measured $p_T^W$ distribution, and the 
${\cal O}(\alpha_s^2)$ 
  predictions for the three choices of scale $\mu=M_W ,~E_T^W ,~p_T^W$;
  with the cuts described in the text.
  b) The ratio of the theoretical predictions, and the preliminary
  \Dzero measurement, to the theoretical prediction with $\mu=M_W$.
Note that in comparing with the \Dzero data, we have integrated over 
the appropriate range of $p_T^W$.
  \label{fig:ptw}}
\end{figure}

The first question we wish to address is the following.  Does the
measured \Dzero $W$ $p_T$ distribution show a similar excess to ${\cal
R}^{10}$ when compared with perturbative QCD predictions?  In figure
\ref{fig:ptw} we show the measured \Dzero $p_T^W$ distribution
\cite{Wpt} for $p_T^W > 20\GeV$ as well as the fixed next-to-leading
order (i.e. ${\cal O}(\alpha_s^2)$) theoretical predictions using the parton
level Monte Carlo program \Dyrad \cite{GGK,GG}.  Experimentally, the
$W$ is tagged in its decay into an electron and neutrino, where the
neutrino is detected though the presence of a significant amount of
missing $E_T$.  To match onto the experimental analysis, we apply the
cuts,
\begin{equation}
E_T^e > 25\GeV, \qquad  |\eta^e| < 1.1,  \qquad \slE_T > 25\GeV,
\end{equation}
to trigger on the event.  Jets are defined with a conesize $\Delta
R_{\rm jet} = 0.7$ and may lie in the rapidity range, $|\eta_{\rm
jet}| < 3.5 $ \cite{HMPC,Jaethesis}.  To simulate the experimental jet algorithm, we cluster
all pairs of partons that lie within $R_{\rm sep}\Delta R_{\rm jet}$ of each
other to form a proto jet, then test that all clustered partons lie
within $\Delta R_{\rm jet}$ of the proto jet \cite{Rsep}.
As a default
parameter, we set $R_{\rm sep} = 1.3$.  
The jet direction and transverse energy is constructed 
using the \Dzero recombination procedure \cite{D0recom}.
The electron must also be
isolated from significant amounts of hadronic energy and we simulate
the isolation criteria by imposing,
\begin{equation}
\Delta R (e,\mbox{jet}) > 0.4 \qquad {\rm for} \qquad E_T^{\rm jet} > 10 \GeV.
\end{equation}
Throughout  we use the CTEQ4M parton densities 
\cite{CTEQ4} with $\alpha_s(M_Z)=0.116$ unless stated otherwise
and set the 
renormalisation scale, at which $\alpha_s$ is evaluated, equal to the
factorisation scale, at which the parton densities are evaluated. For
this joint scale we show $1/\sigma \,  d\sigma / d p_T^W$ for the three
choices $\mu = M_W,~ E_T^W$ and $p_T^W$ corresponding to a medium,
hard and soft scale\footnote{We systematically calculate the total
cross-section, $\sigma$, at next-to-leading order (i.e. ${\cal
O}(\alpha_s)$) and with a scale $\mu=M_W$.}.  We see that the
theoretical prediction is only mildly sensitive to the choice of
scale, which gives us confidence in the prediction for $1/\sigma \, 
d\sigma / d p_T^W$.  Furthermore, it is clear that the ${\cal
O}(\alpha_s^2)$ prediction gives an adequate description of the
preliminary \Dzero data.  This is surprising for two reasons.  First,
the effect of smearing either the jet $E_T$ or $p_T^W$ is not taken
into account. Such smearing can either arise from multiple soft gluon
emission
\cite{ly,ak}\footnote{At the large transverse momenta considered here, 
the effect of soft gluons is roughly to increase the boson transverse
momentum by ${\cal O}(2~\GeV)$ \cite{ak}, which is much less than the
bin width ($5~\GeV$) used in the theoretical calculation.}, or from
experimental mismeasurement of the energy of hadrons.  Second, given
the excess of the observed ${\cal R}^{10}$ ratio over the
\Dyrad predictions, we would naively have expected the theoretical 
predictions to fall short 
for the $W$ transverse momentum distribution as well.

\begin{figure}
\begin{center}
\psfig{figure=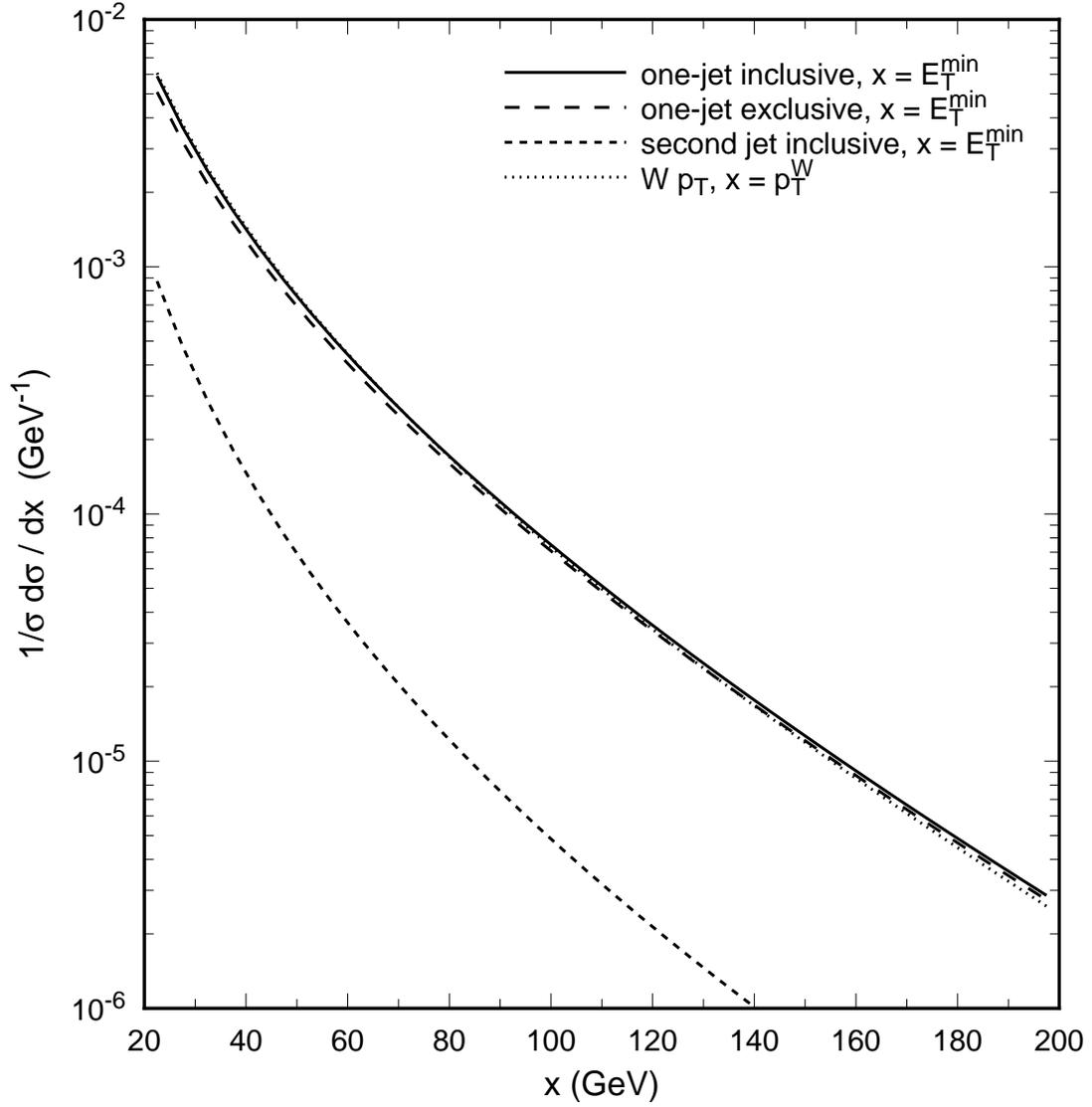,height=16cm}
\end{center}
\caption{The transverse energy distributions for the one-jet inclusive (solid), 
one-jet exclusive (long dashed) and second-jet inclusive (short dashed) rates,
together with the $W$ boson transverse momentum distribution (dotted).
The joint renormalisation/factorisation scale is $\mu = M_W$.
  \label{fig:dsdx}}
\end{figure}

Of course, the ${\cal R}^{10}$ ratio is directly related to the
one-jet exclusive $E_T$ distribution rather than the $W$ boson
transverse momentum distribution.  We show both of these quantities in
Fig.~\ref{fig:dsdx}.  For illustrative purposes, we also show the
one-jet inclusive and second-jet $E_T$ distributions.  As we might
expect, we see that the inclusive one-jet transverse energy
distribution is very similar to the inclusive $W$ transverse
momentum distribution.  At leading order, these two distributions are
identical and this relationship is largely preserved at next-to-leading order.
The one-jet exclusive distribution is suppressed relative to the
one-jet inclusive rate because in some events the second parton is
identified as a jet.  This difference is described by the $E_T$
distribution of the second jet (short dashed line).  From this plot,
we anticipate that from the theoretical point of view, the ${\cal
R}^{10}$ ratio constructed from the single jet exclusive distribution
will be somewhat smaller than the ${\cal R}^{W}$ ratio determined from
the $W$ boson $p_T$ distribution for given $E_T^{\rm min} = p_T^{W,\rm
min}$ values less than around 120~GeV.  As mentioned earlier, in the
theoretical \Dyrad prediction the effect of soft gluon radiation or
hadron mismeasurements on either the jet $E_T$ or the $W$ $p_T^W$ is
not taken into account. Such smearing will generally lead to an
increase in the jet $E_T$ and $W$ $p_T^W$ and hence on both the ${\cal
R}^{W}$ and ${\cal R}^{10}$ ratios. However because ${\cal R}^{W}$ is
sensitive to measurements on all hadrons/gluons in the scattering,
whilst ${\cal R}^{10}$ is only affected by hadrons/gluons that lie
within the jet, we might expect the effect of smearing to increase the
${\cal R}^{W}$ distribution more than the ${\cal R}^{10}$
distribution.

\begin{figure}[t]
\begin{center}
\psfig{figure=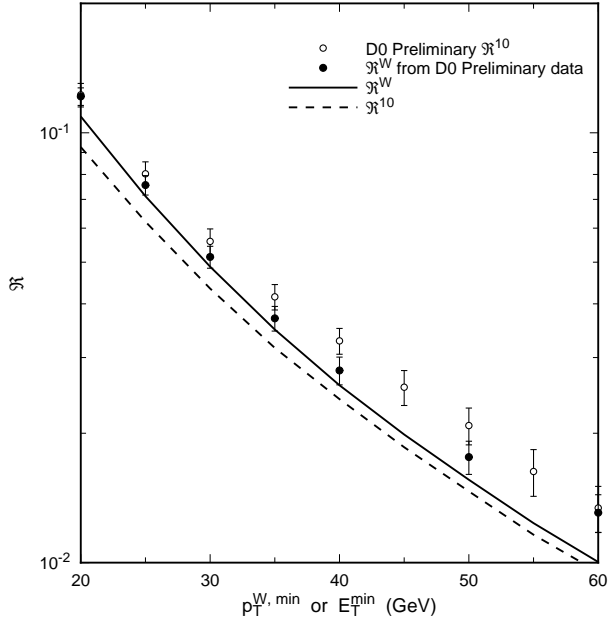,height=9cm}
\end{center}
\caption{Experimental measurements for ${\cal R}^{W}$ and ${\cal R}^{10}$, 
and the corresponding next-to-leading order QCD predictions. 
For the theoretical calculations we have chosen the scale $\mu=M_W$.
\label{fig:RWvsR10}}
\end{figure}

In order to make the difference between the preliminary \Dzero
measurements for ${\cal R}^{10}$ and $1/\sigma \,  d\sigma / d p_T^W$ more
transparent, we convert the measured transverse momentum distribution
into ${\cal R}^{W}$ using equation \ref{eq:ptwtorw}. To estimate the
experimental errors on ${\cal R}^{W}$ we make the assumption that
there are no common systematic errors and merely add the errors in the
$1/\sigma \,  d\sigma / d p_T^W$ distribution in quadrature. We show the
extracted ${\cal R}^{W}$ data in Fig.~\ref{fig:RWvsR10}, as well as
the \Dzero measurement for ${\cal R}^{10}$ and the next-to-leading
order predictions for both quantities.  Note that in extracting ${\cal
R}^{10}$, events may only contribute to the $W$ + 1 jet cross section
when there is exactly one jet observed with $E_T$ above $E_T^{\rm
min}$. Similarly, the $W$ + 0 jet cross section is constructed from
events with no jets observed with an $E_T$ greater than $E_T^{\rm
min}$. In fact, the denominators for both ${\cal R}^{10}$ and
${\cal R}^{W}$ are identical at next-to-leading order.  As
expected from our discussion of the raw $W$ transverse momentum
distribution, the theoretical prediction for ${\cal R}^{W}$ is
consistent with the extracted experimental value, although possibly
slightly low because of the cumulative effect of events at higher $p_T^W$.
We also see that, for a given $E_T^{\rm min} =
p_T^{W,\rm min}$, the measured ${\cal R}^{10}$ value lies above the
experimental ${\cal R}^{W}$ value.  On the other hand, the
corresponding next-to-leading order prediction for ${\cal R}^{10}$
systematically lies below the ${\cal R}^{W}$ value.  This is the
observed discrepancy between the measured ${\cal R}^{10}$ value and
the \Dyrad prediction. In the case of the theoretical prediction the
difference between ${\cal R}^{10}$ and ${\cal R}^{W}$ can be traced
back to the fact that the numerator of ${\cal R}^{10}$ in the \Dzero
definition \cite{HMPC} is the one-jet exclusive rate.  While the
experimental uncertainties in ${\cal R}^{10}$ and ${\cal R}^{W}$ are
somewhat different, the exclusive nature of ${\cal R}^{10}$ ensures
that the next-to-leading order predictions will lie beneath those of
${\cal R}^{W}$ for $E_T^{\rm min} = p_T^{W,\rm min} \leq 100$~GeV. 
Increasing the gluonic content of the proton
or adding new
heavy objects \cite{chou} will affect both ${\cal R}^{10}$ and
${\cal R}^{W}$ but will not change this hierarchy.  
We note in passing that
if the numerator of ${\cal R}^{10}$ is replaced with the $W$ +~1 jet
inclusive rate, then the \Dyrad predictions for ${\cal R}^{10}$ and
${\cal R}^{W}$ are almost identical.

\begin{figure}[t]
\begin{center}
\psfig{figure=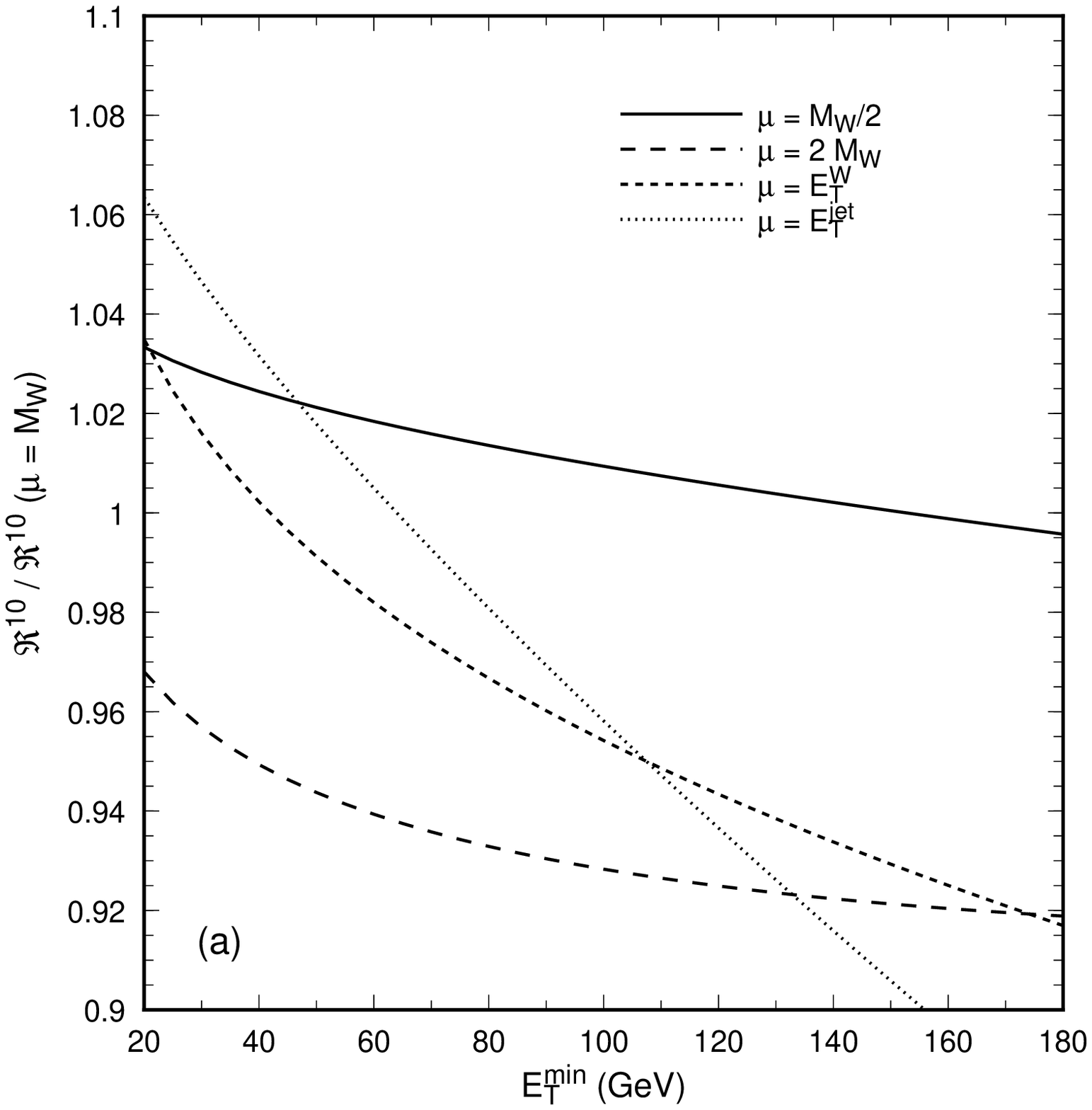,height=8cm}
\psfig{figure=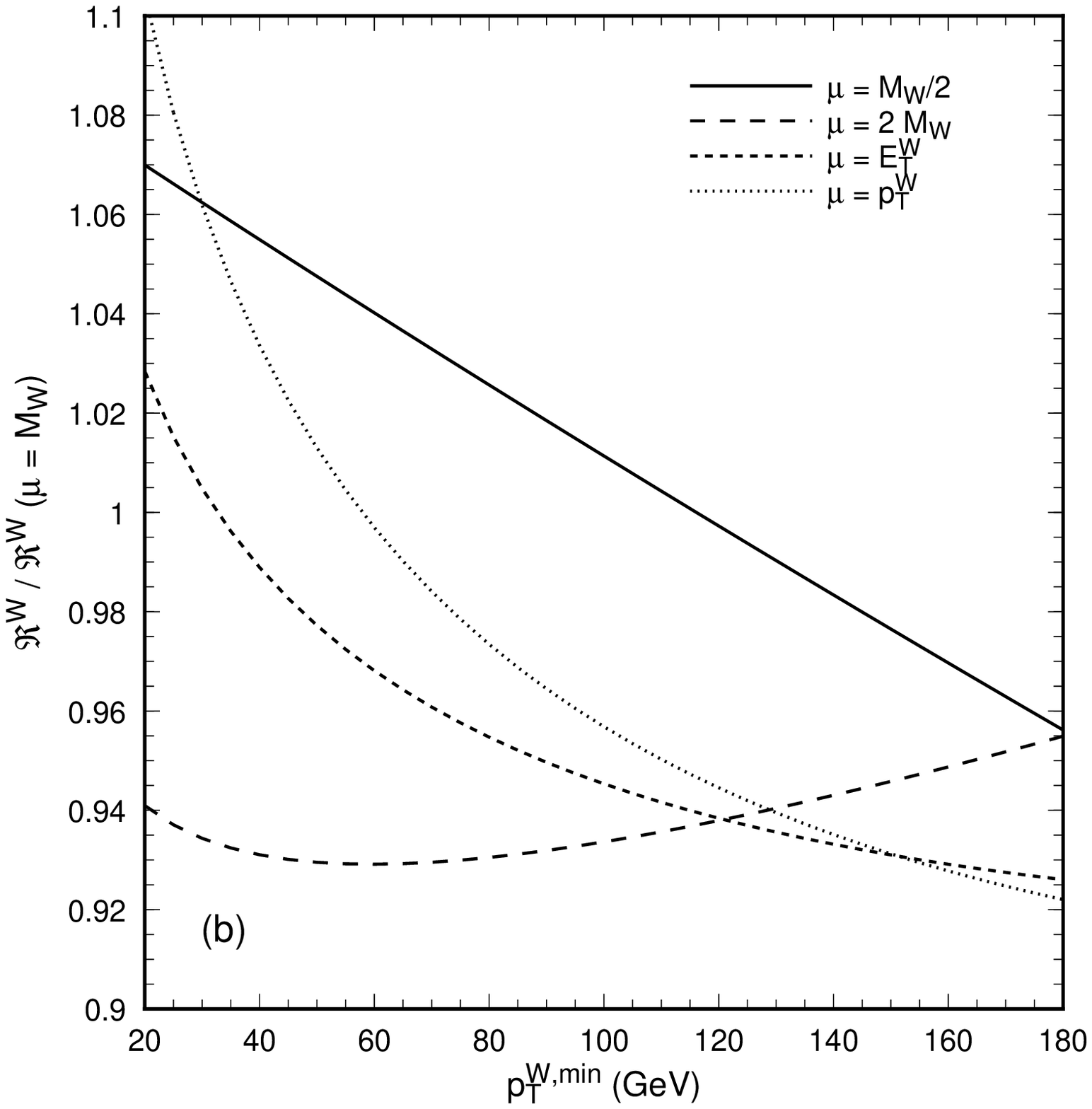,height=8cm}
\end{center}
\caption{The scale dependence for the theoretical prediction for 
(a) ${\cal R}^{10}$ and (b) ${\cal R}^{W}$ relative to that for $\mu = M_W$. 
  \label{fig:R10scale}}
\end{figure}

We now turn to the question of how reliable the predictions for ${\cal
R}^{10}$ and ${\cal R}^{W}$ are under variations of the
renormalisation/factorisation scale.  In Fig.~\ref{fig:R10scale} we
show the relative change compared to $\mu = M_W$ for a soft scale,
$\mu =E_T^{\rm jet}$ or $\mu = p_T^W$ respectively, a hard scale, $\mu
= E_T^W$ and simple multiples of moderate scales $\mu = 0.5 M_W$ and
$\mu = 2M_W$.  We see that changes of $\pm 10\%$ are possible which
are clearly not enough to account for the difference between the
\Dyrad prediction for ${\cal R}^{10}$ and the \Dzero measurement.
Furthermore, both ${\cal R}^{10}$ and ${\cal R}^{W}$ have a similar
behaviour under these scale changes, and the difference between ${\cal
R}^{10}$ and ${\cal R}^{W}$ is relatively insensitive to the changes
of scales.

\begin{figure}[t]
\begin{center}
\psfig{figure=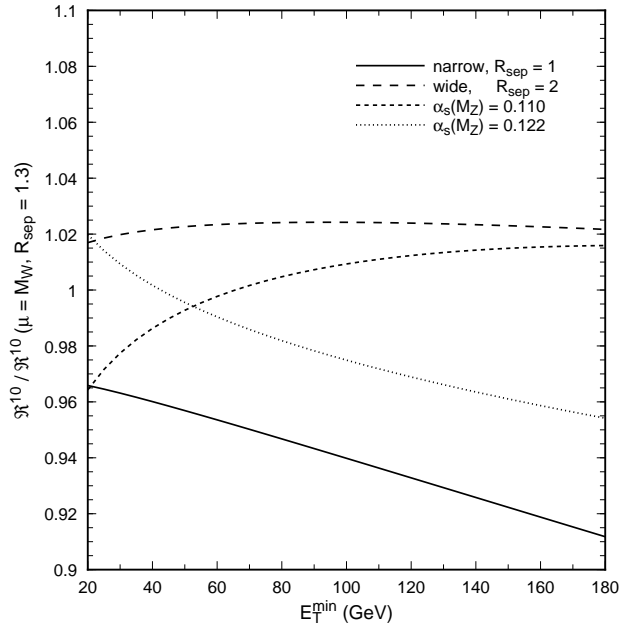,height=9cm}
\end{center}
\caption{The dependence of the \Dyrad prediction for ${\cal R}^{10}$ 
on the jet
  clustering algorithm and on $\alpha_s$. We show the predictions
  normalised to that for $R_{\rm sep}$ clustering and 
  $\alpha_s(M_Z) = 0.116$. In all cases we have chosen $\mu = M_W$.
  \label{fig:R10clus-as}}
\end{figure}

Given that the prediction for ${\cal R}^{W}$ does fit the experimental
measurement so well, we should search for the ways in which the ${\cal
R}^{10}$ measurement differs from the ${\cal R}^{W}$ measurement. The
only major difference between ${\cal R}^{10}$ and ${\cal R}^{W}$ is
that for the ${\cal R}^{10}$ measurement experimental jets must be
formed, whereas this is not necessary for ${\cal R}^{W}$. As \Dyrad
gives a next-to-leading order prediction for ${\cal R}^{10}$ it
includes configurations where two partons get clustered into a single
jet, and this gives it a non trivial dependence on the jet definition.
Experimentally jets are formed from many hadrons, whereas inside
\Dyrad jets are made from at most two partons, and the difference
between these can lead to a mismatch between jet algorithms at the
hadron and parton levels.  For example an experiment might choose to
cluster all hadrons within some $\eta$-$\phi$ distance $\Delta R_{\rm jet}$ from
the centre of a jet.  However, if we theoretically only cluster
partons that lie within an $\eta$-$\phi$ distance of $\Delta R_{\rm jet}$ of
each other then the theoretical jet will never contain partons up to a
distance $\Delta R_{\rm jet}$ away from the centre of the jet. On the other
hand, we could cluster {\em all} pairs of partons to form a proto jet,
and then test if the partons clustered lie inside a distance of
$\Delta R_{\rm jet}$ of the proto jet direction for the proto jet to remain a
genuine jet. This again is far from the experimental procedure, as
there is no ``seed tower'' in the direction around which the proto jet
is first formed. In order to understand this difference between the
experimental and theoretical definition of a jet we use two extreme
jet clustering algorithms in addition to the default algorithm
described earlier with $R_{\rm sep} = 1.3$.
\begin{enumerate}
\item{} {\bf Narrow:} Cluster two partons if they lie within some $\Delta R_{\rm jet}$
  of each other corresponding to $R_{\rm sep} = 1$.
\item{} {\bf Wide:} Cluster all pairs of partons to form a proto jet and then
  test that all clustered partons lie within some $\Delta R_{\rm jet}$ of the
  proto jet direction corresponding to $R_{\rm sep} = 2$.
\end{enumerate}
The narrow and wide jet clustering definitions correspond to the
narrowest and widest theoretical implementation of the experimental
jet definition.  In figure \ref{fig:R10clus-as} we show the dependence
of the ${\cal R}^{10}$ prediction on the parton level jet clustering.
Different jet clusterings vary the theoretical prediction by only a
few percent which is certainly not enough to be able to explain the
measured ${\cal R}^{10}$ excess. The ${\cal R}^{W}$ ratio shows very
little dependence on the type of jet clustering chosen, arising only
from the isolation cut on the observed electron.

Finally we show the dependence of the ${\cal R}^{10}$ ratio on the
value of $\alpha_s$ used in figure \ref{fig:R10clus-as}.  As \Dzero
have noted \cite{R10old} the dependence is less than
expected due to a cancellation between the $\alpha_s$ dependence of
the parton distributions and the hard scattering. Here we only note in
addition that the uncertainty coming from the experimental definition
of a jet is typically similar to the variation in the rate coming from
$\alpha_s$, and so without an accurate understanding of exactly how to
model jet clustering we do not expect measurements of ${\cal R}^{10}$
to be useful in constraining $\alpha_s$. In this sense we feel that
measurements of ${\cal R}^{W}$ (or better still the analogous ratio
from $Z$ boson events) would be a more reliable theoretical method for
measuring $\alpha_s$ as it is largely free from the ambiguities of
defining jets.

In summary, we have found good agreement between the $W$ boson
$p_T$ distribution as reported by \Dzero and fixed order perturbative
calculations.  On the other hand, we find no way to reconcile the
${\cal R}^{10}$ ratio measured by the \Dzero collaboration with the
same theoretical calculation.  Furthermore, because the ${\cal
R}^{10}$ ratio is exclusive in the number of jets, we see that for a
given value of $E_T^{\rm min} = p_T^{W,\rm min}$, the predicted value
for ${\cal R}^{10}$ always lies beneath that for ${\cal R}^{W}$ in the
currently measured range.  This
appears to be in contradiction with the preliminary experimental
results.

\section*{Acknowledgements}
One of us (EWNG) would like to thank the CERN Theory Group 
for their kind hospitality while this work was carried out.

\end{document}